\documentclass[10pt]{article}%\documentstyle[12pt]{article}
\input epsf.sty
\usepackage{mathrsfs}
\usepackage{amsmath}
\usepackage{mathrsfs}
\usepackage{amssymb}
\usepackage{color}
\usepackage{psfrag}
\newcommand{\bea}{\begin{eqnarray}}
\newcommand{\eea}{\end{eqnarray}}
\newcommand{\be}{\begin{equation}}
\newcommand{\ee}{\end{equation}}
\newcommand{\vs}[1]{\vspace{#1 mm}}

\newcommand{\dsl}{\pa \kern-0.5em /}

\newcommand{\pa}{\partial}

\newcommand{\Q}{{\tilde Q}}
\begin{document}
\topmargin 0mm
\oddsidemargin 0mm

\begin{flushright}

USTC-ICTS-12-01\\

\end{flushright}

\vspace{2mm}

\begin{center}

{\Large \bf One loop integrals reduction }

\vs{10}

{\large Yi Sun$^{a,b)}$\footnote{E-mail: sunyi@mail.ustc.edu.cn}, Hao-Ran Chang
$^{a)}$\footnote{E-mail: hrchang@mail.ustc.edu.cn} }

\vspace{4mm}

{\em

 $^a$ Department of Modern Physics,
\\ University of Science and Technology of China,
 Hefei, 230026, Anhui, P. R. China\\

\vs{4}

 $^b$ Interdisciplinary Center for Theoretical Study,
\\University of Science and Technology of China, Hefei, 230026, Anhui, P. R. China\\}

\end{center}

\vs{10}

\begin{abstract}

By further examining the symmetry of external momenta and masses in Feynman integrals, we fulfilled the method proposed by Battistel and Dallabona, and showed that recursion relations in this method can be applied to simplify Feynman integrals directly.
\end{abstract}
\newpage
\section{Introduction}

Feynman diagrams provide a systematical and elegant procedure for the calculation in perturbative quantum field theory. When the calculation goes to high order, for instance, one-loop order, to get more accurate description by many precision experiments, one will
unavoidly face the computation of Feynman integrals from loop diagrams. A general Feynman integral is hard to reach the analytical result directly. To make life easy, one may simplify it by reduction. In Ref.~\cite{pa1}, a pioneering work in one-loop reduction has been carried out by Passarino and Veltman, where tensor integrals can be reduced to basic scalar integrals, then a series of work \cite{pa11, pa2, pa3, pa4} on one-loop, even two-loop \cite{pa5, pa6} reduction, was done and many other approaches were derived. Conventional methods were to reduce the complicated numerator as a function of loop momentum in the Feynman integrals directly. An alternative way is also interesting.

In Ref.~\cite{pa7}, Battistel and Dallabona showed that one could first introduce the Feynman parameters and accomplish the integrals over loop momentum without
any reduction, and then Feynman integrals were left to integrate over Feynman parameters  with complicated numerator (compared with other tensor reduction schemes, e.g., Ref.\cite{pa11}, one can find no more terms here are introduced even though we introduce the Feynman parameters). After studying the symmetry in different momenta in the Feynman integral, which originates from the symmetry of Feynman parameters, they presented a strategy of reducing n-point functions with arbitrary numerator and equal masses to integrals with numerator 1 (hereafter, we denote those integrals as \emph{quasi-basic integrals}), the number of which is up to the power of Feynman parameters in the numerator. The quasi-basic integrals is difficult to deal with analytically with this method. In Ref.\cite{pa7}, it was also mentioned that those quasi-basic integrals could be reduced to basic integrals which could be calculated analytically and were studied very well in Ref.~\cite{pa8, pa9, pa10}. But they did not give the explicit ways to realize it. We examined the symmetry of external momenta and masses in Feynman integrals further and found this method works very well compared with other tensor integrals reduction schemes, e.g., \cite{pa11}. The reason is that terms in recursion relations usually appear in pairs in real calculation considering the property of the introduction of Feynman parameters, therefore the recursion relations can be directly applied to simplify Feynman integrals. This will dramatically simplify the calculation.

Considering such advantage, the method was fulfilled in this work. Firstly, Feynman integrals with different masses are presented since the internal line particles are usually of unequal masses in real calculation, which can be directly applied to massless internal line cases. We secondly show that the recursion relations can be applied to simplify Feynman integrals directly which can lead to dramatic simplification of real calculation. The explicit way to reduce quasi-basic integrals to the corresponding basic integrals was also investigated.

This paper is organized as followed. In Sect.~2, the symmetries in Feynman integrals are studied, and some recursion relations for two-, three-, four-point integrals are presented and the details of the strategy for reducing n-point integrals are also given. In Sect.~3, the recursion relations are applied to simplify Feynman integrals. In Sect.~4, we reduce the quasi-basic integrals to three well-known integrals and discuss the general case. In Sect.~5, the conclusions and discussions are presented.

\section{Recursion relations in Feynman integrals}

In this section, following the method in Ref.~\cite{pa7}, we examine the symmetries of different external momenta and masses of internal lines in Feynman integrals,
 and some useful recursion relations for two-, three- and four-point integrals will be presented. The details
of the strategy of reducing n-point integrals are also studied. With these relations, we can reduce general
integrals with arbitrary numerator to finite quasi-basic integrals in one loop integrals. We first introduce Feynman
parameters and perform the integrals over loop momentum without any reduction, which leaves scalar
integrals over Feynman parameters. The equal masses case can be found in Ref.~\cite{pa7}

\subsection{Two-point function}

The finite part of one loop two-point scalar integral can always be expressed as
\begin{eqnarray}\label{eq1}
Z_{k}(p_1^2,m_1^2;m_0^2)&=&\int^1_0{x^k \ln{\frac{Q(p_1^2,m_1^2,m_0^2)}{\mu^2}}}dx, \\
Q(p_1^2,m_1^2;m_0^2)&=&p_1^2 x (x-1)+m_1^2 x +(1-x)m_0^2 .
\end{eqnarray}

In the above definition, $p_1$ is a momentum carried by an external line or a combination of them, $m_0$ and $m_1$ are masses carried by the propagators, $\mu$ is a parameter with the dimension of mass, which plays the role of a common scale for all the involved physical quantities. We suppose $p_1^2\neq0$ first. For simplicity, we write such integral in a compact way
\begin{eqnarray}
Z_k(i,j)&=&\int_0^1{x^k \ln{\frac{Q(i,j)}{\mu^2}}}dx,\\
Q(i,j)&=&p_i^2 x^2-k_{i,j} x +m_j^2,
\end{eqnarray}
with $k_{i,j}=p_i^2-m_i^2 + m_j^2$.

In the next subsection we will confront some similar structures with more complicated parameters
\begin{eqnarray}
{Z}_{k}^{'\lambda}(i,j)&=&\int_0^1{x^k {Q'}(i,j)^{\lambda}(\ln{\frac{Q'(i,j)}{\mu^2}}}-\sum_{s=1}^{\lambda}{\frac{1}{s}})dx,
\\Q'(i,j)&=&(p_i-p_j)^2 x^2-k_{ij} x +m_j^2,
\end{eqnarray}
with $k_{ij}=(p_i-p_j)^2 -m_i^2 + m_j^2$.

For the cases with arbitrary subscript $k$ in Eq.~(\ref{eq1}), we can use some recursion  relations to reduce them to the cases with subscript 0 instead of integrating them directly. The first such recursion relations are
\begin{eqnarray}
Z_{1}(1,0)&=&\frac{k_{1,0}}{2\,p_1^2}Z_0(1,0)+\frac{m_1^2}{2\, p_1^2}\ln{\frac{m_1^2}{\mu^2}}-\frac{m_0^2}{2\, p_1^2}\ln{\frac{m_0^2}{\mu^2}}+\frac{m_0^2-m_1^2}{2\, p_1^2},
\\
Z_{2}(1,0)&=&\frac{2\,k_{1,0}}{3\,p_1^2}Z_1(1,0)-\frac{m_0^2}{3 p_1^2}Z_0(1,0)-\frac{2}{9}
+\frac{m_1^2}{3\, p_1^2}\ln{\frac{m_1^2}{\mu^2}}+\frac{k_{1,0}}{6\, p_1^2},
\\
Z_{3}(1,0)&=&\frac{3\,k_{1,0}}{4\,p_1^2}Z_2(1,0)-\frac{m_0^2}{2 p_1^2}Z_1(1,0)-\frac{1}{8}
+\frac{m_1^2}{4\, p_1^2}\ln{\frac{m_1^2}{\mu^2}}+\frac{k_{1,0}}{12\, p_1^2},
\\
Z_{4}(1,0)&=&\frac{2\,k_{1,0}}{5\,p_1^2}Z_3(1,0)-\frac{3\,m_0^2}{5 p_1^2}Z_2(1,0)-\frac{2}{25}
+\frac{m_1^2}{5\, p_1^2}\ln{\frac{m_1^2}{\mu^2}}+\frac{k_{1,0}}{20\, p_1^2}.
\end{eqnarray}
Generally, for $n>1$, we have
\begin{eqnarray}
Z_{n}(1,0)&=&\frac{2}{n+1}(\frac{n\,k_{1,0}}{2\,p_1^2}Z_{n-1}(1,0)-\frac{(n-1)\,m_0^2}{2 p_1^2}Z_{n-2}(1,0)-\frac{1}{n+1}\nonumber\\&&+\frac{m_1^2}{2\, p_1^2}\ln{\frac{m_1^2}{\mu^2}}+\frac{k_{1,0}}{2\,n\, p_1^2}).
\end{eqnarray}

With these recursion relations, all the two-point functions can be reduced to \,$Z_0(1,0)$. This also works for massless internal particle cases and the extension is straightforward. The reduction of $Z_{k}^\lambda$ follows the same step.

If $p_1^2=0$ and $m_1^2\neq m_0^2$, we have
\begin{flalign}
&Z_{n}(1,0)=\frac{1}{n+1}(-\frac{1}{n-1}+\frac{m_1^2}{m_1^2-m_0^2}
\ln{\frac{m_1^2}{\mu^2}}-\frac{m_0^2}{n(m_1^2-m_0^2)}Z_{n-1}).
\end{flalign}

If $p_1^2=0$ and $m_1^2=m_0^2$, one can integrate those functions directly and obtain
\begin{flalign}
&Z_{n}(1,0)=\frac{1}{n+1}\ln{\frac{m_0^2}{\mu^2}}.
\end{flalign}

\subsection{Three-point function}

The three-point integrals have richer structures, which can be expressed as
\begin{eqnarray}
\xi_{mn}(p_1,m_1;p_2,m_2;m_0)&=&\int_0^1 dx_1\int_0^{1-x_1}dx_2, \frac{x_1^m\,x_2^n}{Q(p_1,m_1;p_2,m_2;m_0)},\label{eq9}\\
\eta_{mn}(p_1,m_1;p_2,m_2;m_0)&=&\int_0^1 dx_1\int_0^{1-x_1}dx_2 x_1^m\,x_2^n\ln{\frac{Q(p_1,m_1;p_2,m_2;m_0)}{\mu^2}},\label{eq10}
\\
Q(p_1,m_1;p_2,m_2;m_0)&=&p_1^2\,x_1^2+p_2^2\,x_2^2+2\,p_1\cdot p_2\,x_1\,x_2-k_{1,0}\,x_1-k_{2,0}\,x_2+m_0^2.
\end{eqnarray}
For convenience, we introduce some useful compact notations,
\begin{eqnarray}
\xi_{mn}(i,j,k)&=&\int_0^1 dx_1\int_0^{1-x_1}dx_2 \frac{x_1^m\,x_2^n}{Q(i,j,k)},\\
\eta_{mn}^{\lambda}(i,j,k)&=&\int_0^1 dx_1\int_0^{1-x_1}dx_2 x_1^m\,x_2^n\,Q^{\lambda}(\ln{\frac{Q(i,j,k)}{\mu^2}}-\sum_{s=1}^n{\frac{1}{s}}),
\\
Q(i,j,k)&=&p_i^2\,x_i^2+p_j^2\,x_j^2+2\,p_i\cdot  p_j\,x_1\,x_2-k_{i,k}\,x_1-k_{j,k}\,x_2+m_k^2.
\end{eqnarray}
As mentioned in the previous subsection, we will confront some similar structures with more complicated parameters later,
\begin{eqnarray}
\xi'_{mn}(i,j,k)&=&\int_0^1 dx_1\int_0^{1-x_1}dx_2 \frac{x_1^m\,x_2^n}{Q'(i,j,k)},\\
{\eta'}_{mn}^{\lambda}(i,j,k)&=&\int_0^1 dx_1\int_0^{1-x_1}dx_2 x_1^m\,x_2^n\,Q'^{\lambda}(\ln{\frac{Q'(i,j,k)}{\mu^2}}-\sum_{k=1}^n{\frac{1}{k}}),
\\
Q'(i,j,k)&=&(p_i-p_k)^2\,x_i^2+(p_j-p_k)^2\,x_j^2+2\,(p_i-p_k)\cdot (p_i-p_k)\,x_1\,x_2\nonumber\\&&-k_{ik}\,x_1-k_{jk}\,x_2+m_k^2.
\end{eqnarray}
Since $\{p_i,m_i\}$ and $\{p_j,m_j\}$ are symmetric in Eqs.~(\ref{eq9}) and (\ref{eq10}), such integrals are invariant if we exchange the momenta, masses and Feynman parameters  simultaneously, more specifically,
~$ \xi_{mn}(i,j,k)=\xi_{nm}(j,i,k)$~and $ \eta_{mn}(i,j,k)=\eta_{nm}(j,i,k)$~, which is proved very useful in obtaining recursion relations among different integrals.

After integration by part, some recursion relations are presented below.

(i):~$n+m=1$~
\begin{eqnarray}
p_1\cdot p_2\xi_{10}+p_2^2\xi_{01}=\frac{k_{2,0}}{2}\xi_{00}+\frac{1}{2}Z_0^{'}(1,2)-\frac{1}{2}Z_0(1,0),\label{eq18}
\end{eqnarray}
where $\xi_{mn}$ and $\eta_{mn}$ denote $\xi_{mn}(1,2,0)$ and $\eta_{mn}(1,2,0)$, respectively. Such simplification is also used in the case for four-point. With the symmetry mentioned above, we can obtain another two recursion relations by exchanging 1 and 2. As an example, we give the counterpart recursion relation
\begin{eqnarray}
p_1\cdot p_2\xi_{01}+p_1^2\xi_{10}=\frac{k_{1,0}}{2}\xi_{00}+\frac{1}{2}Z_0^{'}(2,1)-\frac{1}{2}Z_0(2,0).\label{eq20}
\end{eqnarray}

By solving  Eqs.~(\ref{eq18}) and (\ref{eq20}), we can reduce the functions \, $\xi_{01}$ and $\xi_{10}$\ to \,$\xi_{00}$, which reads
\begin{eqnarray}
((p_1\cdot p_2)^2-p_1^2p_2^2)\xi_{01}&=&(\frac{k_{1,0}}{2} p_1\cdot p_2-\frac{k_{2,0}}{2} p_1^2)\xi_{00}-\frac{p_1^2}{2}(Z_0(12,2)-Z_0(1,0))\nonumber\\&&+\frac{p_1\cdot p_2}{2}(Z_0(21,1)-Z_0(2,0)),
\\
((p_1\cdot p_2)^2-p_1^2p_2^2)\xi_{10}&=&(-\frac{k_{1,0}}{2} p_2^2+\frac{k_{2,0}}{2} p_1p_2)\xi_{00}+\frac{p_1\cdot p_2}{2}(Z_0(12,2)-Z_0(1,0))\nonumber\\&&-\frac{p_2^2}{2}(Z_0(21,1)-Z_0(2,0)).\label{res}
\end{eqnarray}

The symmetry between ~$\xi_{01}$~ and~$\xi_{10}$~is also obvious in the above expression. The recursion relations between $\eta$ functions have the parallel structures.

If $(p_1\cdot p_2)^2-p_1^2p_2^2=0$, one cannot obtain $\xi_{01}$ and $\xi_{10}$ from Eq.(\ref{res}) any more. But the right-hand side of this equation is equal to zero,which implies that we can express $\xi_{00}$ as a function of $Z_0$ and the explicit form of $\xi_{01}$ can be obtained later in a similar way.

In real calculation, terms in the left-hand side of Eq.(\ref{eq18}) usually appear in pairs considering the property of the introduction of Feynman parameters. This will enable the recursion relation Eq.(\ref{eq18}) to be directly applied, therefore in some cases one does not need to obtain the explicit form as  Eq.(\ref{res}). This can dramatically simplify the calculation and we will explore this simplification in the next section.

Some other recursion relations are given as\\
(ii) :~$n+m=2$~
\begin{flalign}
&p_1\cdot p_2\xi_{20}+p_2^2\xi_{11}=\frac{k_{2,0}}{2}\xi_{10}
+\frac{1}{2}Z_1^{'}(1,2)-\frac{1}{2}Z_1(1,0),
\\
&p_1\cdot p_2\xi_{11}+p_1^2\xi_{20}=
\frac{k_{1,0}}{2}\xi_{10}-\frac{1}{2}\eta_{00}
-\frac{1}{2}Z_1^{'}(2,1)+\frac{1}{2}Z_0^{'}(2,1).\label{eq26}
\end{flalign}

Another two similar recursion relations can be obtained by exchanging 1 and 2, specifically, ${p_1,m_1}$ and ${p_2,m_2}$ in the above recursion relations. Then $\xi_{20}$, $\xi_{11}$ and $\xi_{02}$ can be expressed as a function of $\xi_{00}$.

If $(p_1\cdot p_2)^2-p_1^2p_2^2=0$, by the similar way as done in the case for $n+m=1$, we can express $\xi_{10}$ as a function of $\eta_{00}$ and $Z$ functions. The explicit form of $\xi_{11}$ can be obtained similarly. We will give the general rule for the arbitrary-point function.

(iii):~$n+m=3$~
\begin{flalign}
&p_1\cdot p_2\xi_{30}+p_2^2\xi_{21}=\frac{k_{2,0}}{2}\xi_{20}+\frac{1}{2}Z_2^{'}(1,2)-\frac{1}{2}Z_2(1,0),
\nonumber\\
&p_1\cdot p_2\xi_{12}+p_2^2\xi_{03}=\frac{k_{2,0}}{2}\xi_{02}-\eta_{01}+\frac{1}{2}Z_2^{'}(1,2)-Z_1^{'}(1,2)+\frac{1}{2}Z_0^{'}(1,2),
\nonumber\\
&p_1\cdot p_2\xi_{21}+p_2^2\xi_{12}=\frac{k_{2,0}}{2}\xi_{11}-\frac{1}{2}\eta_{10}-\frac{1}{2}Z_2^{'}(1,2)+\frac{1}{2}Z_1^{'}(1,2).
\end{flalign}

It can be found that the symmetry between different momenta and masses provides us with a new way of reducing Feynman integrals. Integrals with structures Eqs.~(\ref{eq9}) and (\ref{eq10}) can be reduced to two quasi-basic integrals $\xi_{00}$, $\eta_{00}$ and $Z_{0}$. In fact, only $\xi_{00}$ and $Z_{0}$ are basic integrals as will be performed in Sec.4, which means what we should do is just to deal with the basic integrals.

\subsection{Four-point function}

In the same way, the four-point integrals can be expressed as
\begin{eqnarray}
\zeta_{mnl}&=&\zeta_{mnl}(1,2,3,0)=\xi_{mnl}(p_1,m_1;p_2,m_2;p_3,m_3;m_0)
\nonumber\\&=&\int_0^1 dx_1\int_0^{1-x_1}dx_2\int_0^{1-x_1-x_2}dx_3 \frac{x_1^m\,x_2^n\,x_3^l}{(Q(p_1,m_1;p_2,m_2;p_3,m_3;m_0))^2},
\\\xi_{mnl}&=&\xi_{mnl}(1,2,3,0)=\xi_{mnl}(p_1,m_1;p_2,m_2;p_3,m_3;m_0)
\nonumber\\&=&\int_0^1 dx_1\int_0^{1-x_1}dx_2\int_0^{1-x_1-x_2}dx_3\, \frac{x_1^m\,x_2^n\,x_3^l}{Q(p_1,m_1;p_2,m_2;p_3,m_3;m_0)},
\\
\eta_{mnl}&=&\eta_{mnl}(1,2,3,0)=\eta_{mnl}(p_1,m_1;p_2,m_2;p_3,m_3;m_0)
\nonumber\\&=&\int_0^1 dx_1\int_0^{1-x_1}dx_2\int_0^{1-x_1-x_2}dx_3 \, x_1^m\,x_2^n\,x_3^l\ln{\frac{Q(p_1,m_1;p_2,m_2;p_3,m_3;m_0)}{\mu^2}},
\end{eqnarray}
\begin{eqnarray}
Q(p_1,m_1;p_2,m_2;p_3,m_3;m_0)&=&p_1^2\,x_1^2+p_2^2\,x_2^2+p_3^2\,x_3^2+2\,p_1\cdot p_2\,x_1\,x_2
+2\,p_1\cdot p_3\,x_1\,x_3
\nonumber\\&&+2\,p_2\cdot p_3\,x_2\,x_3
-k_{1,0}\,x_1-k_{2,0}\,x_2-k_{3,0}\,x_3+m_0^2.
\end{eqnarray}

As in the case for three-point integrals, the symmetry and recursion relations are also exit, which will help to reduce the above general integrals to quasi-basic integrals and some three- and two-point integrals. Considering the symmetry between momenta and masses in the above definition, we have
\begin{flalign} &\zeta_{mnl}(i,j,k,l)=\zeta_{mln}(i,k,j,l)=\zeta_{lnm}(k,j,i,l)=\zeta_{nml}(j,i,k,l)=\zeta_{lmn}(k,i,j,l)=\zeta_{nlm}(j,k,i,l),~
\nonumber\\ &\xi_{mnl}(i,j,k,l)=\xi_{mln}(i,k,j,l)=\xi_{lnm}(k,j,i,l)=\xi_{nml}(j,i,k,l)=\xi_{lmn}(k,i,j,l)=\xi_{nlm}(j,k,i,l),~
\nonumber\\ &\eta_{mnl}(i,j,k,l)=\eta_{mln}(i,k,j,l)=\eta_{lnm}(k,j,i,l)=\eta_{nml}(j,i,k,l)=\eta_{lmn}(k,i,j,l)=\eta_{nlm}(j,k,i,l).~
\end{flalign}

(i) :~$n+m+l=1$~
\begin{flalign}
&p_1\cdot p_3 \zeta_{100}+p_2p_3\zeta_{010}+p_3^2\zeta_{001}=\frac{k_{3,0}}{2}\zeta_{000}-\frac{1}{2}\xi_{00}^{'}+\frac{1}{2}\xi_{00},
\nonumber\\
&p_1\cdot p_3 \xi_{100}+p_2p_3\xi_{010}+p_3^2\xi_{001} =\frac{k_{3,0}}{2}\xi_{000}+\frac{1}{2}\eta_{00}^{'}-\frac{1}{2}\eta_{00}.
\end{flalign}

Another four similar recursion relations can be obtained by exchanging 1, 2 and 3. These six recursion relations can reduce the integrals $\zeta_{001}$, $\zeta_{010}$, $\zeta_{100}$, $\xi_{001}$, $\xi_{010}$ and $\xi_{100}$ to $\zeta_{000}$ and $\xi_{000}$.

This method will not work any more if the determinant is equal to zero, i.e.
\begin{equation}
\det \left( \begin{array}{ccc}
p_1^2 & p_1\cdot p_2 & p_1\cdot p_3  \\
%\multicolumn{6}{c}\hrulefill \\
p_1\cdot p_2 & p_2^2 & p_2\cdot p_3  \\
p_1\cdot p_3 & p_2\cdot p_3 & p_3^2  \\
 \end{array} \right)=0.
\end{equation}

We will give a general rule in the arbitrary-point function for this special case.

(ii) :~$n+m+l=2$~
\begin{flalign}
&p_1\cdot p_3 \zeta_{200}+p_2\cdot p_3\zeta_{110}+p_3^2\zeta_{101} =\frac{k_{3,0}}{2}\zeta_{100}-\frac{1}{2}\xi_{10}^{'}+\frac{1}{2}\xi_{10},
\nonumber\\
&p_1\cdot p_3 \zeta_{110}+p_2\cdot p_3\zeta_{020}+p_3^2\zeta_{011} =\frac{k_{3,0}}{2}\zeta_{010}-\frac{1}{2}\xi_{01}^{'}+\frac{1}{2}\xi_{01},
\nonumber\\
&p_1\cdot p_3 \zeta_{101}+p_2\cdot p_3\zeta_{011}+p_3^2\zeta_{002} =\frac{k_{3,0}}{2}\zeta_{001}+\frac{1}{2}\xi_{000}-\frac{1}{2}(\xi'_{00}-\xi'_{10}-\xi'_{01}),
\end{flalign}
with~$\xi'_{ij}=\xi_{ij}^{'}(1,2,3)$~and ~$\eta'_{ij}=\eta_{ij}^{'}(1,2,3)$~. By exchanging momenta and masses, nine different recursion relations are obtained. But, as expected, only six of them are independent, corresponding to six functions in the case for $n+m+l=2$. Similar structures are found for the functions $\xi'$s. For every three integrals, three recursion relations group a closed set of equations, so it is easy to solve them.

\begin{flalign}
&p_1\cdot p_3 \xi_{200}+p_2\cdot p_3\xi_{110}+p_3^2\xi_{101} =\frac{k_{3,0}}{2}\xi_{100}+\frac{1}{2}\eta_{10}^{'}-\frac{1}{2}\eta_{10},
\nonumber\\
&p_1\cdot p_3 \xi_{110}+p_2\cdot p_3\xi_{020}+p_3^2\xi_{011} =\frac{k_{3,0}}{2}\xi_{010}+\frac{1}{2}\eta_{01}^{'}-\frac{1}{2}\eta_{01},
\nonumber\\
&p_1\cdot p_3 \xi_{101}+p_2\cdot p_3\xi_{011}+p_3^2\xi_{002}=\frac{k_{3,0}}{2}\xi_{001}-\frac{1}{2}\eta_{000}+\frac{1}{2}(\eta'_{00}-\eta'_{10}-\eta'_{01}).
\end{flalign}

\subsection{The arbitrary-point function}

In this subsection, the details of reducing the arbitrary-point integrals are presented. As will be proved later, it is possible to reduce the arbitrary n-point functions with arbitrary numerator and different masses to quasi-basic integrals. This also holds for massless cases and the extension is straightforward.

The general arbitrary k-point integrals have structures as
\begin{flalign}
&\xi_{i_1,i_2,\cdot\cdot\cdot, i_k}^n =\xi_{i_1,i_2,...,i_k}^n(1,2,\cdot\cdot\cdot,k,0)=\xi_{i_1,i_2,\cdot\cdot\cdot, i_k}^n(p_1,m_1;p_2,m_2;\cdot\cdot\cdot ;p_k,m_k;m_{0})\nonumber\\&=\int_0^1 dx_1dx_2\cdot\cdot\cdot dx_k \frac{x_1^{i_1}\,x_2^{i_2}\cdot\cdot\cdot\,x_k^{i_k}}{Q^n},\label{eq60}\\
&\eta_{i_1,i_2,\cdot\cdot\cdot ,i_k}^n =\eta_{i_1,i_2,...,i_k}^n(1,2,\cdot\cdot\cdot,k,0)=\eta_{i_1,i_2,\cdot\cdot\cdot ,i_k}^n(p_1,m_1;p_2,m_2;\cdot\cdot\cdot ;p_k,m_k;m_{0})\nonumber\\&=\int_0^1 dx_1dx_2\cdot\cdot\cdot dx_k \,x_1^{i_1}\,x_2^{i_2}\cdot\cdot\cdot\,x_k^{i_k}Q^n
(\ln{\frac{Q}{\mu^2}}-\sum_{s=1}^n{\frac{1}{s}}),\label{eq61}\\
&Q=\sum_{i=1}^n (p_i^2x_i^2-k_{i,0}x_i)+\sum_{0<i<j<n+1}2p_{i}\cdot p_{j}x_{i}x_{j}+m_0^2.
\end{flalign}

After integration by parts, the general recursion relations read

for $ n\neq 1 $
\begin{eqnarray}
&&p_1\cdot p_k\xi_{i_1+1,i_2,\cdot\cdot\cdot ,i_k}^n + p_2\cdot p_k\xi_{i_1,i_2+1 ,i_3,\cdot\cdot\cdot ,i_k}^n+\cdot\cdot\cdot+p_k^2\xi_{i_1,\cdot\cdot\cdot ,i_{k-1},i_{k}+1}^n\nonumber\\
%&=&\int_0^1 dx_1dx_2\cdot\cdot\cdot dx_{k-1} \frac{x_1^{i_1}\,x_2^{i_2}\cdot\cdot\cdot\,x_k^{i_k}}{2 \,Q^n}dQ+\frac {k_{k,0}}{2}\xi_{i_1,i_2,\cdot\cdot\cdot ,i_k}^n\nonumber\\
%&=&\int_0^1 dx_1dx_2\cdot\cdot\cdot dx_{k}\frac{\partial}{\partial x_k}\,\frac{x_1^{i_1}\,x_2^{i_2}\cdot\cdot\cdot\,x_k^{i_k}}{2 (1-n)\,Q^{n-1}}
%+\frac{i_k}{2(n-1)}\xi_{i_1,i_2,\cdot\cdot\cdot ,i_k-1}^{n-1}+\frac{k_{k,0}}{2}\xi_{i_1,i_2,\cdot\cdot\cdot ,i_k}^n\nonumber\\
&=&\frac{1}{2(1-n)}\Bigl[\sum_{j_1+j_2+\cdot\cdot\cdot+j_k=i_k} (-1)^{i_k-j_k}\,\frac{i_k!}{j_1!j_2!\cdot\cdot\cdot j_k!}\xi_{i_1+j_1,i_2+j_2,\cdot\cdot\cdot,i_{k-1}+j_{k-1}}^{n-1}-(1-sign(i_k))\xi_{i_1,i_2,\cdot\cdot\cdot,i_{k-1}}^{n-1}\Bigl]\nonumber\\
&&+\frac{i_k}{2(n-1)}\xi_{i_1,i_2,\cdot\cdot\cdot ,i_k-1}^{n-1}+\frac{k_{k,0}}{2}\xi_{i_1,i_2,\cdot\cdot\cdot ,i_k}^n\\
&=&\mathcal{I}_{k}^{n},\nonumber\end{eqnarray}

for ~$n=1$~
\begin{eqnarray}\label{eq64}
\mathcal{I}_k^{1}&=&\frac{1}{2}\Bigl[\sum_{j_1+j_2+\cdot\cdot\cdot+j_k=i_k} (-1)^{i_k-j_k}\,\frac{i_k!}{j_1!j_2!\cdot\cdot\cdot j_k!}(\eta_{i_1+j_1,i_2+j_2,\cdot\cdot\cdot,i_{k-1}+j_{k-1}}^{k,1}-(1-sign(i_k))\eta_{i_1,i_2,\cdot\cdot\cdot,i_{k-1}}\Bigl]
\nonumber\\&&
-\frac{i_k}{2}\eta_{i_1,i_2,\cdot\cdot\cdot ,i_k-1}+\frac{k_{k,0}}{2}\xi_{i_1,i_2,\cdot\cdot\cdot ,i_k}.
\end{eqnarray}

To be clear, extra commas were introduced above. Exchanging the $k$-th momentum and mass with others, we can get k-1 independent recursion relations.

Together with Eq.~(\ref{eq64}), we have enough independent equations to solve the corresponding k integrals. Solving these equations, we have,
\begin{eqnarray}
\det(C)\xi^n_{i_1,i_2,\cdot\cdot\cdot i_j+1,\cdot\cdot\cdot
,i_k}=\sum_{l=1}^{k}A_{jl}\mathcal{I}_l^n,
\end{eqnarray}
with
\begin{equation}
C=
\left( \begin{array}{ccccc}
p_1^2 & p_1\cdot p_2 & p_1\cdot p_3 & \cdots & p_1\cdot p_k \\
%\multicolumn{6}{c}\hrulefill \\
p_1\cdot p_2 & p_2^2 & p_2\cdot p_3 & \cdots & p_2\cdot p_k \\
p_1\cdot p_3 & p_2\cdot p_3 & p_3^2 & \cdots &  p_3\cdot p_k\\
\vdots & \vdots & \vdots & \ddots & \vdots \\
p_1\cdot p_k & p_2\cdot p_k & p_3\cdot p_k & \cdots & p_k^2
\end{array} \right),\label{arb}
\end{equation}
and $A_{jl}$ is the algebraic cofactor of $C_{jl}$.

If $\det(C)=0$, we have k equations, \begin{equation}\sum_{l=1}^{k}A_{jl}\mathcal{I}^n_l=0.\end{equation}

As in the case for three-point function, we can obtain the explicit form of $\xi_{i_1,i_2 ,i_3,\cdot\cdot\cdot ,i_k}^n$.

The recursion relations between integrals as Eq.~(\ref{eq61}) have similar structures,

\begin{eqnarray}
&&p_1p_k\eta_{i_1+1,i_2,\cdot\cdot\cdot ,i_k}^n + p_2p_k\eta_{i_1,i_2+1 ,i_3,\cdot\cdot\cdot ,i_k}^n+\cdot\cdot\cdot+p_k^2\eta_{i_1,\cdot\cdot\cdot ,i_{k-1},i_{k}+1}^n
\nonumber\\
%&=&\int_0^1 dx_1dx_2\cdot\cdot\cdot dx_{k-1} \frac{1}{2}x_1^{i_1}\,x_2^{i_2}\cdot\cdot\cdot\,x_k^{i_k}d(\frac{Q^{n+1}}{n+1}(\ln{\frac{Q}{\mu^2}}-\sum_{k=1}^n{\frac{1}{k}}))+\frac {k_{k,0}}{2}\eta_{i_1,i_2,\cdot\cdot\cdot ,i_k}^n\nonumber\\
&=&\frac{1}{2(n+1)}\Bigl[\sum_{j_1+j_2+\cdot\cdot\cdot+j_k=i_k} (-1)^{i_k-j_k}\,\frac{i_k!}{j_1!j_2!\cdot\cdot\cdot j_k!}\eta_{i_1+j_1,i_2+j_2,\cdot\cdot\cdot,i_{k-1}+j_{k-1}}^{k,n+1}-(1-sign(i_k))\eta_{i_1,i_2,\cdot\cdot\cdot,i_{k-1}}^{n+1}\Bigl]\nonumber\\
&&-\frac{i_k}{2(n+1)}\eta_{i_1,i_2,\cdot\cdot\cdot ,i_k-1}^{n+1}+\frac {k_{k,0}}{2}\eta_{i_1,i_2,\cdot\cdot\cdot ,i_k}^n,
\end{eqnarray}
with
\begin{eqnarray}
\xi_{i_1,i_2,\cdot\cdot\cdot ,i_{k-1}}^{\lambda,n}&=&\xi_{i_1,i_2,\cdot\cdot\cdot ,i_{k-1}}^{'n}(1,2,\cdot\cdot\cdot,(\lambda-1),(\lambda+1),\cdot\cdot\cdot,k ,\lambda)~, \\
~\eta_{i_1,i_2,\cdot\cdot\cdot ,i_{k-1}}^{\lambda,n}&=&\eta_{i_1,i_2,\cdot\cdot\cdot,i_{k-1}}^{'n}(1,2,\cdot\cdot\cdot,(\lambda-1),(\lambda+1),\cdot\cdot\cdot,k ,\lambda).~
\end{eqnarray}

The case for the determinant which is equal to zero can be dealt with in a parallel way.

Step by step, the arbitrary n-point functions with arbitrary numerator and different masses can be expressed in terms of the quasi-basic integrals. More quasi-basic integrals are needed for the higher power of Feynman parameters in the numerator, so it is necessary to reduce these quasi-basic integrals to the well-studied integrals if we try to compute them analytically.

In real calculation, terms in the left-hand side of the above recursion relations  usually accompany with each other which will enable the recursion relations here to be directly applied, therefore the calculation will be dramatically simplified since one does not need to obtain the explicit forms of every functions.

\section{Recursion relations applied to Feynman integrals}

In our work, terms in the left-hand side of the recursion relations given in the previous section usually appear in pairs in real calculation considering
the property of the introduction of Feynman parameters. Hence the recursion relations can be applied to simplify the Feynman integrals which will lead to a dramatical simplification of the calculation, while the method in Ref.\cite{pa11} has no such advantage.

For 3-point function, in the scheme of dimensional regularization, Feynman integrals have structures as
\begin{flalign}
I^3_{\mu}=\mu^{4-D}\int \frac{d^D k}{(2\pi)^D}\frac{2k_{\mu}}{A_0A_1A_2},
\end{flalign}
with $A_i=(k+p_i)^2-m_i^2+i\varepsilon$ and $p_0=0$. This equation can be expressed in terms of the functions we defined in the previous section,
\begin{flalign}
i (4\pi)^2 I^3_{\mu}=-p_{1\mu}\xi_{10}-p_{2\mu}\xi_{01}.
\end{flalign}

The explicit forms of $I^3_{\mu_1}$, $I^3_{\mu_1\mu_2}$, $I^3_{\mu_1\mu_2\mu_3}$ can be found in Ref.\cite{pa7}. We find simplification is available when the bare indices are contracted by loop mentum or external momentum which are frequently encountered in real calculation, e.g.
\begin{flalign}
i(4\pi)^2 p_1^{\mu}I^3_{\mu}=-p_1^2\xi_{10}-p_1\cdot p_2\xi_{01}.
\end{flalign}

The right-hand side of the above equation is exactly the same as the left-hand side of Eq.(\ref{eq20}) and the recursion relation there can be directly used. It is not necessary to obtain the explicit forms of both $\xi_{01}$ and $\xi_{10}$ as in Eq.(\ref{res}). We give some nontrivial cases in the following.
\begin{flalign}
&i(4\pi)^2 p_1^{\nu}I^3_{\mu\nu}=\frac{1}{2}p_1^{\mu}\eta_{00}+p_1^{\mu}(p_1^2\xi_{20}+p_1\cdot p_2\xi_{11})
+p_2^{\mu}(p_1^2\xi_{11}+p_1\cdot p_2\xi_{02})
+d.t. \nonumber\\
&=p_1^{\mu}(\frac{k_{1,0}}{2}\xi_{10}-\frac{1}{2}Z_1^{'}(2,1)
+\frac{1}{2}Z_0^{'}(2,1))
+p_2^{\mu}(\frac{k_{1,0}}{2}\xi_{01}+\frac{1}{2}Z_1^{'}(2,1)
-\frac{1}{2}Z_1(2,0))+d.t.,
\end{flalign}
where d.t. denotes the divergent terms in the modified minimal substraction scheme which is proportional to $\frac{2}{4-D}-\gamma+\log(4\pi)$. The treatment of such terms are available in Ref.\cite{pa7} and hence are not considered here.

As we showed above, the recursion relations Eq.(\ref{eq26}) given in the previous section can be applied directly to the right-hand side of the above equation, therefore the
most complicated terms are simplified. This can dramatically simplify the calculation.

It is also worthy to mention that such simplifications holds for the general cases, e.g. the internal particles are massless or the determinant defined in Eq.(\ref{arb}) is equal to zero.

With more bare indices being contracted, more recursion relations, such as
\begin{flalign}
&i(4\pi)^2 p_1^{\mu}p_1^{\nu}I^3_{\mu\nu}
=\frac{k_{1,0}}{2}(p_1^2\xi_{10}+p_1\cdot p_2\xi_{01})+p_1^2(-\frac{1}{2}Z_1^{'}(2,1)+\frac{1}{2}Z_0^{'}(2,1))
\nonumber\\&+p_1\cdot p_2(\frac{1}{2}Z_1^{'}(2,1)-\frac{1}{2}Z_1(2,0))+d.t.,
\end{flalign}
can be applied directly.

Generally speaking, such simplification
always works since the feature of the introduction of Feynman parameters guarantees that terms in recursion relations appear in pairs. Some other examples are presented as
\begin{eqnarray}
i(4\pi)^2k^{\nu}I^3_{\mu\nu}&=&-3p_1^{\mu}\eta_{10}-3p_2^{\mu}\eta_{01}-\frac{1}{6}(p_1^{\mu}+p_2^{\mu})
-p_1^{\mu}(k_{1,0}\xi_{20}+k_{2,0}\xi_{11}-m_0^2\xi_{10})
\nonumber\\&&-p_2^{\mu}(k_{1,0}\xi_{11}+k_{2,0}\xi_{02}-m_0^2\xi_{01})+d.t.,\nonumber\\
i(4\pi)^2k^{\mu}p_1^{\nu}I^3_{\mu\nu}&=&
-3(p_1^{2}\eta_{10}+p_1\cdot p_2\eta_{01})
-\frac{1}{6}(p_1^2+p_1\cdot p_2)
-k_{1,0}(p_1^2\xi_{20}+p_1\cdot p_2\xi_{11})-\nonumber\\&&k_{2,0}(p_1^2\xi_{11}+p_1\cdot p_2\xi_{20})
+m_0^2(p_1^2\xi_{10}+p_1\cdot p_2\xi_{01})+d.t..
\end{eqnarray}

For Feynman integrals in the four-point case, such simplification can be achieved in a similar way. In the scheme of dimensional regularization, they have structures as
\begin{flalign}
I^4_{\mu}=\mu^{4-D}\int \frac{d^D k}{(2\pi) D}\frac{6 k_{\mu}}{A_0A_1A_2A_3}.
\end{flalign}

This equation can be expressed in terms of the functions defined in the previous section,
\begin{flalign}
&i (4\pi)^2 I^4_{\mu}=p_{1\mu}\zeta_{100}+p_{2\mu}\zeta_{010}+P_{3\mu}\zeta_{001}.
\end{flalign}

One can find the explicitly form of $I^4_{\mu_1}$, $I^4_{\mu_1\mu_2}$, $I^4_{\mu_1\mu_2\mu_3}$ and $I^4_{\mu_1\mu_2\mu_3\mu_4}$ in Ref.\cite{pa7}.

As in the three-point case, recursion relations can be applied directly when the bare indices are contracted by loop mentum or external momentum which are frequently encountered in real calculation. We give some examples here.
\begin{eqnarray}
i (4\pi)^2 p_1^{\mu}I^4_{\mu}&=&p_{1}^2\zeta_{100}+p_1\cdot p_{2}\zeta_{010}+p_1\cdot P_{3}\zeta_{001},\nonumber\\
i(4\pi)^2 p_1^{\nu}I^4_{\mu\nu}&=&\frac{1}{2}p_{1\mu}\xi_{000}-p_{1\mu}(p_1^2\zeta_{200}+p_1\cdot p_2\zeta_{110}+p_1\cdot p_3\zeta_{101})
\nonumber\\
&&-p_{2\mu}(p_1^2\zeta_{110}+p_1\cdot p_2\zeta_{020}+p_1\cdot p_3\zeta_{011})-p_{3\mu}(p_1^2\zeta_{101}+p_1\cdot p_2\zeta_{011}+p_1\cdot p_3\zeta_{002}),\nonumber\\
i(4\pi)^2k^{\nu}I^3_{\mu\nu}&=
&-2(p_{1\mu}\xi_{100}+p_{2\mu}\xi_{010}+p_{3\mu}\xi_{001})
+p_{1\mu}(k_{1,0}\zeta_{200}+k_{2,0}\zeta_{110}+k_{2,0}\zeta_{101}
-m_0^2\zeta_{100})\nonumber\\
&&+p_{2\mu}(k_{1,0}\zeta_{110}+k_{2,0}\zeta_{020}+k_{2,0}\zeta_{011}
-m_0^2\zeta_{010})\nonumber\\&&+p_{3\mu}(k_{1,0}\zeta_{101}+k_{2,0}\zeta_{011}+k_{2,0}\zeta_{002}
-m_0^2\zeta_{001}),\nonumber\\
i(4\pi)^2k^{\mu}p_1^{\nu}I^3_{\mu\nu}&=&
-2(p_{1}^2\xi_{100}+p_1\cdot p_{2}\xi_{010}+p_1\cdot p_{3}\xi_{001})
+k_{1,0}(p_{1}^2\xi_{200}+p_1\cdot p_{2}\xi_{110}+p_1\cdot p_{3}\xi_{101})\nonumber\\&&+k_{2,0}(p_{1}^2\xi_{110}+p_1\cdot p_{2}\xi_{020}+p_1\cdot p_{3}\xi_{111})
+k_{3,0}(p_{1}^2\xi_{101}+p_1\cdot p_{2}\xi_{011}+p_1\cdot p_{3}\xi_{002})\nonumber\\&&-m_0^2(p_{1}^2\xi_{100}+p_1\cdot p_{2}\xi_{010}+p_1\cdot p_{3}\xi_{001}).\label{recursion four}
\end{eqnarray}

For the Feynman integrals in the arbitrary $(n+1)$-point case with $(n>2)$, such simplification is also worthy to consider. In the scheme of dimensional regularization, they have structures as
\begin{flalign}
I^{n+1}_{\mu_1\mu_2\cdot\cdot\cdot\mu_m}=\mu^{4-D}\int \frac{d^D k}{(2\pi)^D}\frac{n(n-1)k_{\mu_1}k_{\mu_2}\cdot\cdot\cdot k_{\mu_m}}{A_0A_1\cdot\cdot\cdot A_n}
\end{flalign}

This equation can be expressed in terms of the functions defined in the previous section,
\begin{flalign}
i(4\pi)^2I^{n+1}_{\mu}=(-1)^{n+1}\sum_{l=1}^{n}p_{l\mu}\xi^{n-1}_{[l]}.
\end{flalign}

The notation $[l]$ in $\xi^{n-1}_{[l]}$ means only the $l$-th subscript in $\xi_{i_1,\cdots,i_l,\cdots, i_k}^{n-1}$ defined in Eq.(\ref{eq60}) is 1 while others are all equal to zero. The recursion relations can be applied directly when the bare indices are contracted. We give some examples here.
\begin{flalign}
&i(4\pi)^2p_1^{\mu}I^{n+1}_{\mu}=(-1)^{n+1}\sum_{i=1}^{n}p_1p_{i}\xi^{n-1}_{[i]},\nonumber\\
&i(4\pi)^2k^{\mu}I^{n+1}_{\mu}=(-1)^{n}(\sum_{i=1}^{n}k_{i,0}\xi^{n+1}_{[i]}
-m^2_0\xi^{n-1}_{0\cdot\cdot\cdot 0}+\frac{n}{n-2}\xi^{n-2}_{0\cdot\cdot\cdot 0}),\nonumber\\
&i(4\pi)^2p_1^{\nu}I^{n+1}_{\mu\nu}=(-1)^{n}
(\sum_{i=1}^{n}p_{i\mu}\sum_{j=1}^{n}p_1\cdot p_{j}\xi^{n-1}_{[i,j]}-\frac{1}{2(n-2)}p_{1\mu}\xi_{0,\cdot\cdot\cdot 0}),\nonumber\\
&i(4\pi)^2k^{\nu}I^{n+1}_{\mu\nu}=(-1)^{n}
(\frac{5-n}{n-2}\sum_{i=1}^{n}p_{i\mu}\xi^{n-2}_{[i]}
-\sum_{i=1}^{n}p_{i\mu}\sum_{j=1}^{n}k_{j,0}\xi^{n-1}_{[i,j]}
+m_0^2\sum_{i=1}^{n}p_{i\mu}\xi^{n-1}_{[i]}),
\nonumber\\
&i(4\pi)^2k^{\mu}p_1^{\nu}I^{n+1}_{\mu\nu}=(-1)^{n}
(\frac{5-n}{n-2}\sum_{j=1}^{n}p_1\cdotp_{i}\xi^{n-2}_{[i]}
-\sum_{i=1}^{n}k_{i,0}\sum_{j=1}^{n}p_1\cdot p_j\xi^{n-1}_{[i,j]}
+m_0^2\sum_{j=1}^{n}p_1\cdot p_{i}\xi^{n-1}_{[i]})
\end{flalign}
If n=3, one gets the explicit expressions exactly as Eq.(\ref{recursion four}) in the four-point case as expected.

Due to the recursion relations, the most complicated terms in the right-hand side of the above equation can be simplified dramatically. Such simplification
always works since terms in recursion relations appear in pairs.

\section{Basic integrals}

After the reduction in Sect~2, some quasi-basic integrals, i.e., $\eta_{0,0,\cdot\cdot\cdot,0}^{n}$ and $\xi_{0,0,\cdot\cdot\cdot,0}^{n}$ are obtained. As has been mentioned in Ref.\cite{pa7}, these quasi-basic integrals can be reduced to the basic integrals which can be calculated analytically. But they did not give the explicit ways to realize it. To fulfill this method, in this section, we show the explicit way to reduce the quasi-basic integrals to the corresponding basic integrals.

With the symmetry in Feynman integrals, one can get such relations in three- and four-point integrals as
\begin{flalign}
 &\eta_{00}=-\frac{1}{2}(k_{1,0} \xi_{10}+k_{2,0} \xi_{01}-2 m_0^2\xi_{00})-\frac{1}{2}
 +\frac{1}{2}(Z_0^{'}(1,2)-Z_1^{'}(1,2)+Z_0^{'}(2,1)-Z_1^{'}(2,1)),
 \\ &\eta_{000}=-\frac{1}{3}(k_{1,0} \xi_{100}+k_{2,0} \xi_{010}+k_{3,0} \xi_{001}-2m_0^2\xi_{000}) -\frac{1}{9}
+\frac{1}{3}(\eta_{00}^{'}(1,2,3)-\eta_{01}^{'}(1,2,3)
-\eta_{10}^{'}(1,2,3)
 +\nonumber\\&\eta_{00}^{'}(1,3,2)-\eta_{01}^{'}(1,3,2)-\eta_{10}^{'}(1,3,2)
 +\eta_{00}^{'}(3,2,1)-\eta_{01}^{'}(3,2,1)-\eta_{10}^{'}(3,2,1)),
 \\
 &\xi_{000}=(k_{1,0} \zeta_{100}+k_{2,0} \zeta_{010} +k_{3,0} \zeta_{001}-2 m_0^2\zeta_{000})+\xi_{00}^{'}(1,2,3)-\xi_{01}^{'}(1,2,3)-\xi_{10}^{'}(1,2,3)
 \nonumber\\&+\xi_{00}^{'}(1,3,2)-\xi_{01}^{'}(1,3,2)-\xi_{10}^{'}(1,3,2) +\xi_{00}^{'}(3,2,1)-\xi_{01}^{'}(3,2,1)-\xi_{10}^{'}(3,2,1).
\end{flalign}

So $\eta_{00}$, $\eta_{000}$ and $\xi_{000}$ can be reduced to the widely studied functions $\zeta_{000}$, $\xi_{00}$ and $Z_{0}$. In fact, every functions we got in Sect.~2 can be reduced, following a parallel step, to basic integrals in the arbitrary point integrals,
 \begin{flalign}
 \eta_{0,0,\cdot\cdot\cdot,0}^n =(\eta_{0,0,\cdot\cdot\cdot,0}^{k,n}-\eta_{0,1,\cdot\cdot\cdot,0}^{k,n}-\cdot\cdot\cdot-\eta_{0,0,\cdot\cdot\cdot,1}^{k,n})
-\int_0^1 dx_1dx_2\cdot\cdot\cdot dx_{k-1}\,x_k (\ln{\frac{Q}{\mu^2}}-\sum_{s=1}^{n-1}{\frac{1}{s}}) dQ^n.
\end{flalign}

As before, we can get k similar relations, which are symmetric by exchanging momenta, masses and Feynman parameters simutaneously. After summing them up, we obtain the general relations,
\begin{flalign}
&(k+2\,n)\eta_{0,0,\cdot\cdot\cdot,0}^n =\sum_{\lambda=1}^{k}(\eta_{0,0,\cdot\cdot\cdot,0}^{\lambda,n}-\eta_{0,1,\cdot\cdot\cdot,0}^{\lambda,n}-\cdot\cdot\cdot-\eta_{0,0,\cdot\cdot\cdot,1}^{\lambda,n})
 \nonumber\\&-n(k_{1,0}\eta_{1,0,\cdot\cdot\cdot,0}^{n-1}+\cdot\cdot\cdot+k_{k,0}\eta_{0,\cdot\cdot\cdot,0,1}^{n-1}-2\,m_0^2\eta_{0,\cdot\cdot\cdot,0,0}^{n-1})
-2\int_0^1 dx_1dx_2\cdot\cdot\cdot dx_k Q^\lambda.
\end{flalign}

Similarly,
\begin{eqnarray}
(k-2\,n)\xi_{0,0,\cdot\cdot\cdot,0}^n &=&\sum_{\lambda=1}^{k}(\xi_{0,0,\cdot\cdot\cdot,0}^{\lambda,n}
-\xi_{0,1,\cdot\cdot\cdot,0}^{\lambda,n}-\cdot\cdot\cdot
-\xi_{0,0,\cdot\cdot\cdot,1}^{\lambda,n})
\nonumber\\&&+n(k_{1,0}\xi_{1,0,\cdot\cdot\cdot,0}^{n+1}+\cdot\cdot\cdot
  +k_{k,0}\xi_{0,\cdot\cdot\cdot,0,1}^{n+1}
  -2\,m_0^2\xi_{0,\cdot\cdot\cdot,0,0}^{n+1}).\label{eq72}
\end{eqnarray}

If $k-2\,n=0$, the left-hand of Eq.(\ref{eq72}) disappears, which means the reduction chain is cut off. If k is odd, corresponding to even-point function, such cutoff will be absent, while k is even, corresponding to odd-point function, the final results for n-point integrals with ~$n>4$~ will need two basic integrals, $\xi^{\frac{(n-1)}{2}}$ and $\xi^{n}$. To be specifically, for two-, three-, four- and $(2\,n+1)$-point integrals, only one corresponding basic integral is needed.

\begin{figure}
\begin{center}
  \includegraphics{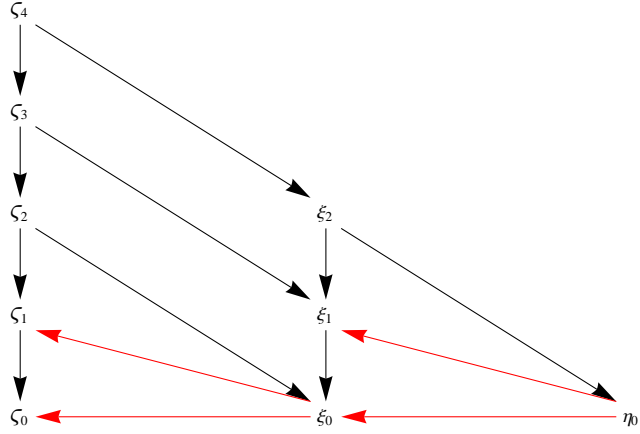}
  \end{center}
  \caption{ $\zeta_{nml}$, $\xi_{nml}$ and $\eta_{nml}$ with $n+m+l=\lambda$ are denoted as $\zeta_{\lambda}$, $\xi_{\lambda}$ and $\eta_{\lambda}$ respectively in the four-point function. $\lambda$ will decrease 2 in every reduction to quasi-basic integrals as the black lines show, but increase only 1 in the reduction to the basic integral according to the red lines.}\label{fig1}
\end{figure}

Another question to which attention should be paid is whether the reduction will be entrapped into a closed loop. The answer is no! In Sect.~2, $i_1+i_2+\cdot\cdot\cdot+i_k$ will decrease 2 in every reduction, but increase only 1 in Sect.~4. We show a simple example explicitly in Fig.(\ref{fig1}), so it is safe to reduce the general Feynman integrals with the method.

\section{Conclusions and discussions}

By taking the advantage of the symmetry of  momenta and masses in the Feynman integrals, one-loop integrals reduction can be carried out explicitly. After one introduces the Feynman parameters and accomplishes the integrals over the loop momentum, Feynman
integrals have the structures as Eqs.~(\ref{eq60}) and (\ref{eq61}). It is found that, originating from the symmetry of Feynman parameters, such integrals are invariant if we exchange the momenta, masses and Feynman parameters simultaneously. In Ref.~\cite{pa7}, with those symmetries, Battistel and Dallabona presented a strategy of reducing n-point functions with arbitrary numerator and equal masses to quasi-basic integrals and argued that those quasi-basic integrals could be reduced to basic integrals.

We examined the symmetry of external momenta and masses in Feynman integrals further and found this method works very well compared with other tensor integrals reduction schemes, e.g., \cite{pa11}. The reason is that terms in the recursion relations usually appear in pairs in real calculation considering the feature of the introduction of Feynman parameters, therefore the recursion relations can be directly applied to simplify the Feynman integrals. Specifically speaking, such simplification is available when loop momentum is contracted, which is frequently encountered in real calculation instead of carrying bare Lorentz index. More recursion relations can be applied if more loop momenta are contracted. This will dramatically simplify the calculation. We also presented the explicit way to reduce the quasi-basic integrals to the corresponding basic integrals and found that in n-point integrals, for n is even, Feynman integral can be reduced to only one basic integral, while n is odd and larger than 4, it can be reduced to two basic integrals.
In the cases for three- or four-point integrals that we are mostly concerned about and frequently encounter, only one basic integral is necessary, which is just the well studied integral in much previous work. The extension of this method to the general case with different masses are also considered. This also works for massless internal particle cases and the extension is straightforward.

Considering the advantage of this method, we will realize a computer program for three-, four- and five-point integrals reduction later.

\section*{Acknowledgements:}
We are grateful to Prof. Dao-Neng Gao for helpful discussion and this work is supported by National Natural Science
Foundation of China (Grant No.11075149 and No.10975128)

%%%% 参考文献排版格式：


\begin{thebibliography}{99}
\itemsep=-4pt plus.2pt minus.2pt  %% 调整参考文献条与条之间的间距
\small
%\bibitem{1}  %% 输入参考文献1内容
%\bibitem{2}  %% 输入参考文献2内容
%\bibitem{3} %% 输入参考文献3内容，其余依次往下排

\bibitem{pa1}
Passarino~G, Veltman~M~J~G. Nucl. Phys. B, 1979, {\bf 160}:151
\\
\bibitem{pa11}
Duplancic~G, Nizic~B. Eur. Phys. J. C, 2004, {\bf 35}:105
\\
\bibitem{pa2}
Denner~A, Dittmaier~S. Nucl. Phys. B, 2006, {\bf 734}:62; Nucl. Phys. B, 2003, {\bf 658}:175
\\
\bibitem{pa3}
Fleischer~J, Riemann~T. Phys. Rev. D, 2011, {\bf 83}:073004
\\
\bibitem{pa4}
Fleischer~J, Jegerlehner~F, Tarasov~O~V. Nucl. Phys. B, 2000, {\bf 566}:423
\\
\bibitem{pa5}
Weiglein~G, Scharf~R, Bohm~M. Nucl. Phys. B, 1994, {\bf 416}:606
\\
\bibitem{pa6}
Anastasiou~C, Gehrmann~T, Oleari~C, Remiddi~E, Tausk~J~B. Nucl. Phys. B, 2000, {\bf 580}:577
\\
\bibitem{pa7}
Battistel~O~A, Dallabona~G. Eur. Phys. J. C, 2006, {\bf 45}:721
\\
\bibitem{pa8}
't Hooft~G, Veltman M~J~G. Nucl. Phys. B, 1979, {\bf 153}:365
\\
\bibitem{pa9}
Denner~A, U.~Nierste~U, Scharf~R. Nucl. Phys. B, 1991, {\bf 367}:637
\\
\bibitem{pa10}
Denner~A, Dittmaier~S. Nucl. Phys. B, 2011, {\bf 844}:199

\end{thebibliography}
\end{document}